\documentclass[journal]{./IEEEtran}

\usepackage{cite}

\ifCLASSINFOpdf
   \usepackage[pdftex]{graphicx}
\else
\fi
\usepackage{amsmath}
\usepackage{algorithmic}
\usepackage[thinlines]{easytable}
\usepackage{multirow}

\begin{document}
%
\title{A Sparse Sampling Sensor Front-end IC for Low Power Continuous SpO$_2$ \& HR Monitoring}
%
%
%
        
\author{Sina Faraji Alamouti, \IEEEmembership{Student Member,~IEEE}, Jasmine Jan, \IEEEmembership{Student Member,~IEEE}, Cem Yalcin, \IEEEmembership{Student Member,~IEEE}, Jonathan Ting, Ana Claudia Arias, \IEEEmembership{Senior Member,~IEEE}, Rikky Muller, \IEEEmembership{Senior Member,~IEEE}
\thanks{S.F. Alamouti, J. Jan, C. Yalcin, J. Ting, A.C.Arias, and R. Muller are with the Department of Electrical Engineering and Computer Sciences, University of California, Berkeley, Berkeley, CA USA 94720}
\thanks{R. Muller is with Chan-Zuckerberg Biohub, San Francisco, CA USA 94158.}}

%
%


\markboth{}%
{Shell \MakeLowercase{\textit{et al.}}: Bare Demo of IEEEtran.cls for IEEE Journals}

%



\maketitle

\begin{abstract}
Photoplethysmography (PPG) is an attractive method to acquire vital signs such as heart rate and blood oxygenation and is frequently used in clinical and at-home settings. Continuous operation of health monitoring devices demands a low power sensor that does not restrict the device battery life. Silicon photodiodes (PD) and LEDs are commonly used as the interface devices in PPG sensors; however, using of flexible organic devices can enhance the sensor conformality and reduce the cost of fabrication. In most PPG sensors, most of system power consumption is concentrated in powering LEDs, traditionally consuming mWs. Using organic devices further increases this power demand since these devices exhibit larger parasitic capacitances and typically need higher drive voltages. This work presents a sensor IC for continuous SpO$_2$ and HR monitoring that features an on-chip reconstruction-free sparse sampling algorithm to reduce the overall system power consumption by $\sim$70\% while maintaining the accuracy of the output information. The designed frontend is compatible with a wide range of devices from silicon PDs to organic PDs with parasitic capacitances up to 10 nF. Implemented in a 40 nm HV CMOS process, the chip occupies 2.43 mm$^2$ and consumes 49.7 µW and 15.2 µW of power in continuous and sparse sampling modes respectively. The performance of the sensor IC has been verified \textit{in vivo} with both types of devices and the results are compared against a clinical grade reference. Less than 1 bpm and 1\% mean absolute errors were achieved in both continuous and sparse modes of operation.  
\end{abstract}


%
\IEEEpeerreviewmaketitle

\section{Introduction}
%
%
%
%

\IEEEPARstart{T}{imely} diagnosis of many chronic cardiovascular and respiratory systems diseases can be enabled through continuous monitoring of vital signs such as heart rate (HR), blood oxygenation level (SpO$_2$), respiration rate, blood pressure etc. \cite{b1,b2,b3,b4,b5,b6}. Abrupt changes in respiration rate and SpO$_2$ can be a marker of serious illness in many pulmonary diseases \cite{b1}. A comfortable, wearable device can track these vital signs autonomously, unobtrusively, and without the user’s intervention, allowing changes to be immediately detected and reported to medical staff, preventing disease progression. Similarly in patients with a history of heart failure, remote monitoring of vital signals has proven essential in early recognition of potential congestions \cite{b2,b6}. Moreover, in the setting of COVID-19, remote monitoring provides the healthcare workforce with real-time biodata without needing any physical contact, reducing the spread of infection \cite{b3}. As a result, the interest and market for biosensors in at home care continue to grow. A similar trend is observed for health monitoring wearable devices including smart-watches, rings, and health patches \cite{b4}. These devices aim to unobtrusively record a user’s vital signs without any need for their intervention. The biodata is recorded, stored, and presented to the user via application interfaces and can be used to warn users of an abnormal condition. Traditionally, electrocardiography (ECG) has been used to accurately measure HR and cardiac waveforms; since it requires access to multiple sites across the body, the subject has to trigger these measurements and thus, it cannot be performed continuously. Photoplethysmography (PPG) is an attractive method to acquire biodata such as HR due to its fully optical, non-invasive nature where the sensors do not even need to contact the subject skin. In addition, using light from two distinct wavelengths, typically Red and infra-red (IR), the SpO$_2$ of the subject can be extracted, a technique known as pulse oximetry. Figure 1 shows the operation of a pulse oximeter where light sources, typically LEDs, are driven sequentially and the reflected light is received by a photodiode (PD), inducing a photocurrent that contains the PPG signal. A current sensing IC, shown in Figure 1, then samples and digitizes the photocurrent to output the information. 
\begin{figure}[!t]
    \setlength\abovecaptionskip{2\baselineskip}
    \centering
    \includegraphics[width=\columnwidth]{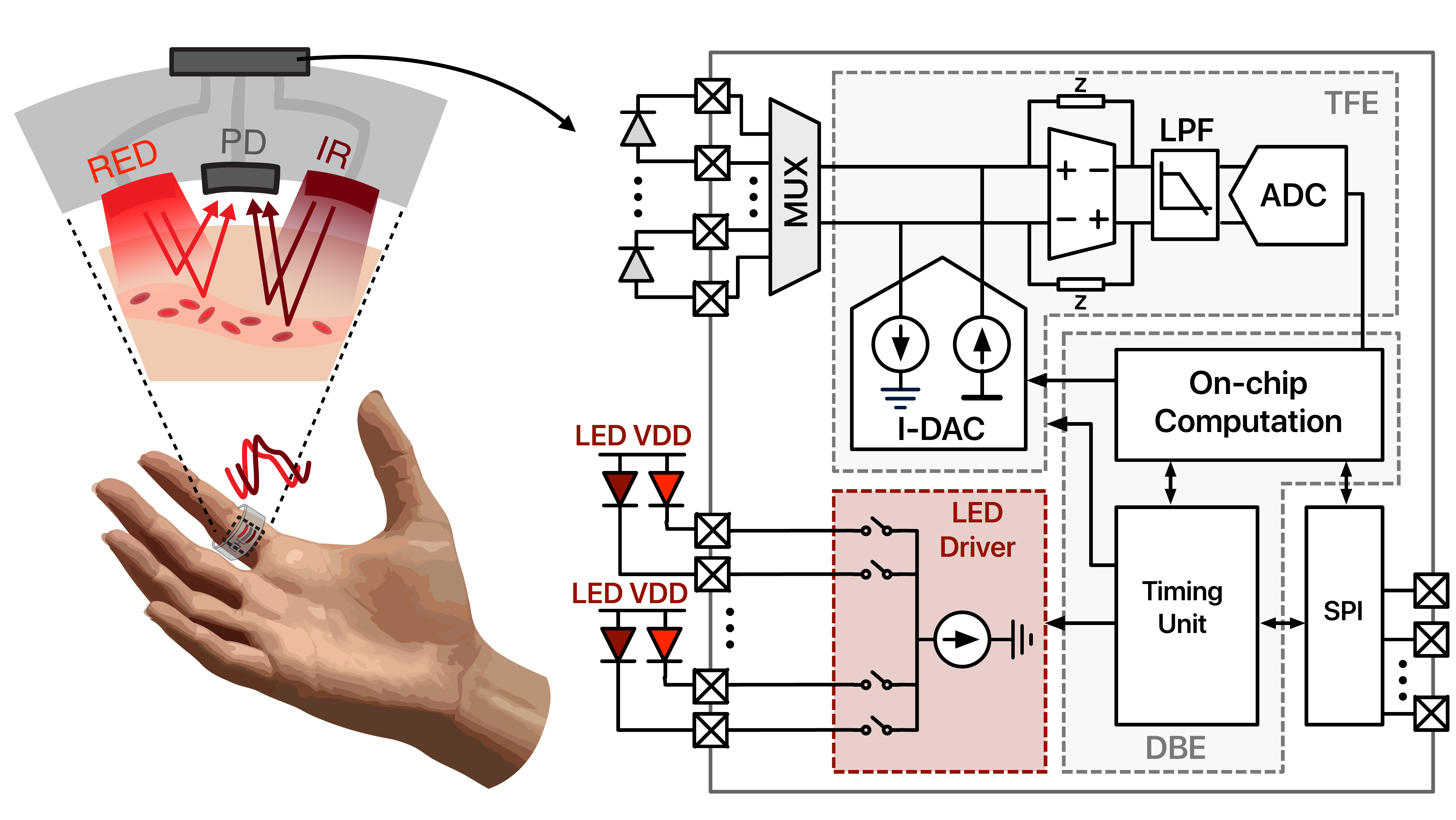}
    \caption{Reflectance mode pulse oximetry and typical SpO$_2$ sensing IC block diagram.}
    \label{fig1}
\end{figure}

Silicon PDs and LEDs are commonly used in today’s SpO$_2$ sensors. Despite great responsivity, efficiency, and mm-scale sizes, they are still rigid devices which eventually impact the dimensions and conformality of the overall sensor. Organic optical devices however are lightweight, mechanically flexible, and shock resistant, and therefore offer a more conformal solution, improving user comfort \cite{b7,b8,b9}. Flexible vital sign monitoring patches can therefore be built to seamlessly integrate into clothing and be comfortably worn over long durations \cite{b9,b10}. In addition, these devices can be printed using low-cost fabrication processes, reducing the overall cost of the health monitoring patches \cite{b11,b12}. 

Commercial SpO$_2$ sensors operate with mWs of power to drive the LEDs posing strict limits on the sensor battery life. Use of OLEDs further restricts the battery life as these devices generally require higher drive voltages, up to 8 V \cite{b9}. Duty cycling has been used in many prior arts \cite{b14,b15} to lower the power consumption of the overall system at the cost of increased noise bandwidth, impacting Signal-to-Noise Ratio (SNR). The parasitic capacitance of the PD (C$\mathrm{_{Par}}$) impacts the input referred noise of the transimpedance amplifier (TIA) by attenuating the feedback factor as the frequency increases \cite{b16}. C$\mathrm{_{Par}}$ in silicon PDs ranges from sub-pF to a few 100s of pF depending on the size of the PD active area. OPDs however tend to exhibit a much larger C$\mathrm{_{Par}}$ up to 10 nF which can easily exceed the maximum capacitance handled by many of the prior arts. It is therefore critical that a sensor IC aimed for a wearable can operate with a wide variety of devices and can handle a wide range C$\mathrm{_{Par}}$.

\cite{b14,b17} employed on-chip photodetectors to significantly lower the detector and the interface parasitic capacitance and as a result compensate the SNR penalties at extremely low duty cycle ratios. However, on-chip photodiodes generally tend to offer inferior responsivities compared to their off-the-shelf counterparts, reducing the power saving benefits of this approach. In addition, there are fundamental limitations on how fast the LEDs can operate as well as the response bandwidth of the PDs, restricting the lower bound of the duty cycle. Moreover, this technique cannot be extended to using organic devices as they are usually fabricated on plastic substrates.  \cite{b18} tried to balance the tradeoff between SNR and the LEDs and readout power by setting the front-end bias current. The technique significantly reduced the readout power, but the LEDs still dominated the sensor power consumption. Compressive sampling was first introduced in \cite{b19} to exploit the sparse nature of the PPG signal and thus save power by reducing the number of sample points. Despite excellent results for HR estimation from the compressively sampled data, SpO$_2$ measurements required full reconstruction of the PPG waveform out of randomly selected samples, demanding up to 10 mW of processing power \cite{b19}. \cite{b20} presented the heart-beat-locked-loop technique to make the sensor lock to the PPG signal period and selectively sample the PPG peaks to report HR data. This lowered the LED power by a factor of $\sim$6.5$\times$, but since the PPG waveform itself was not digitized, no SpO$_2$ measurements were performed.

This work presents an SpO$_2$ and HR monitoring IC utilizing a reconstruction-free sparse sampling algorithm to reduce the overall system power consumption by about 70\%. This manuscript is organized as follows. Section II outlines the system requirements as well as the overall architecture of the sensor. In Section III, the proposed sparse sampling algorithm is discussed. Section IV describes the circuit-level details of the implemented IC. Bench-top electrical and \textit{in vivo} measurement results are presented in section V followed by the conclusions and comparison against selected prior arts in section VI.

\section{System Overview}
\label{sec:overview}
\subsection{System requirements}

As discussed in \cite{b21}, the PD photocurrent contains multiple components including the detector dark current, the ambient photocurrent, and the reflected light’s baseline (I$\mathrm{_{DC}}$) and pulsatile (I$\mathrm{_{AC}}$) components. The overall input baseline component is commonly 40-60 dB stronger than the pulsatile signal. Thus, without any subtraction, the readout chain will need a very large dynamic range, greater than 100 dB \cite{b22}, which can pose challenges in realizing a low power readout.

The PPG I$\mathrm{_{AC}}$ is a low frequency signal with most of its power residing between 0.5 to 5 Hz, which according to the Nyquist theorem can be sampled with frequencies as low as 10 Hz. However, as explained in \cite{b20}, the timing resolution of the PPG samples can impact the accurate detection of pulsatile peaks, which in turn affects the reported HR error of the sensor. Therefore, a sampling frequency of a few 10s to 100s of Hz is typically selected. In this work, to achieve an average HR error of less than 1 bpm, a sampling rate of 100 S/s is used. Furthermore, to extract SpO$_2$ with a 2\% error, the sensor requires nearly 40 dB of SNR \cite{b23}, which determines the noise level of the readout. The amplitude of I$\mathrm{_{AC}}$ depends on many biological factors such as the structure, thickness, and color of Epidermis and Dermis, optical factors such as wavelength of light and devices' responsivities, and mechanical factors such as distance between the sensor and tissue as well as the angle of emission. These factors cause the I$\mathrm{_{AC}}$ to vary over a wide range from a few nAs to up to 100s of nAs. Thus, to maintain the required SNR for low amplitude inputs, the readout input referred noise density needs to be approximately 1-10 pA/rtHz.

\subsection{Sensor architecture}

The block diagram of a typical SpO$_2$ sensing IC is shown in Figure 1. The LEDs are driven in sequential phases using on-chip current sources, emitting light at either R (660 nm) or Green (530nm) as well as IR (880 nm) wavelengths towards the tissue. The received photocurrent with a baseline component as large as a few µAs is then digitized by the readout. To relax the dynamic range requirements of the chain, the large DC component of the input is subtracted by means of differential current DACs (I-DACs) at the input. The remaining AC component is then amplified, filtered, and digitized through the readout chain. On-chip digital back-end computes the required DC code for the DACs, closing the servo-loop. The digitized data can then be transmitted to an FPGA or a PC via a serial programming interface.
\begin{figure}[!t]
    \setlength\abovecaptionskip{-0.3\baselineskip}
    \centering{\includegraphics[width=\columnwidth]{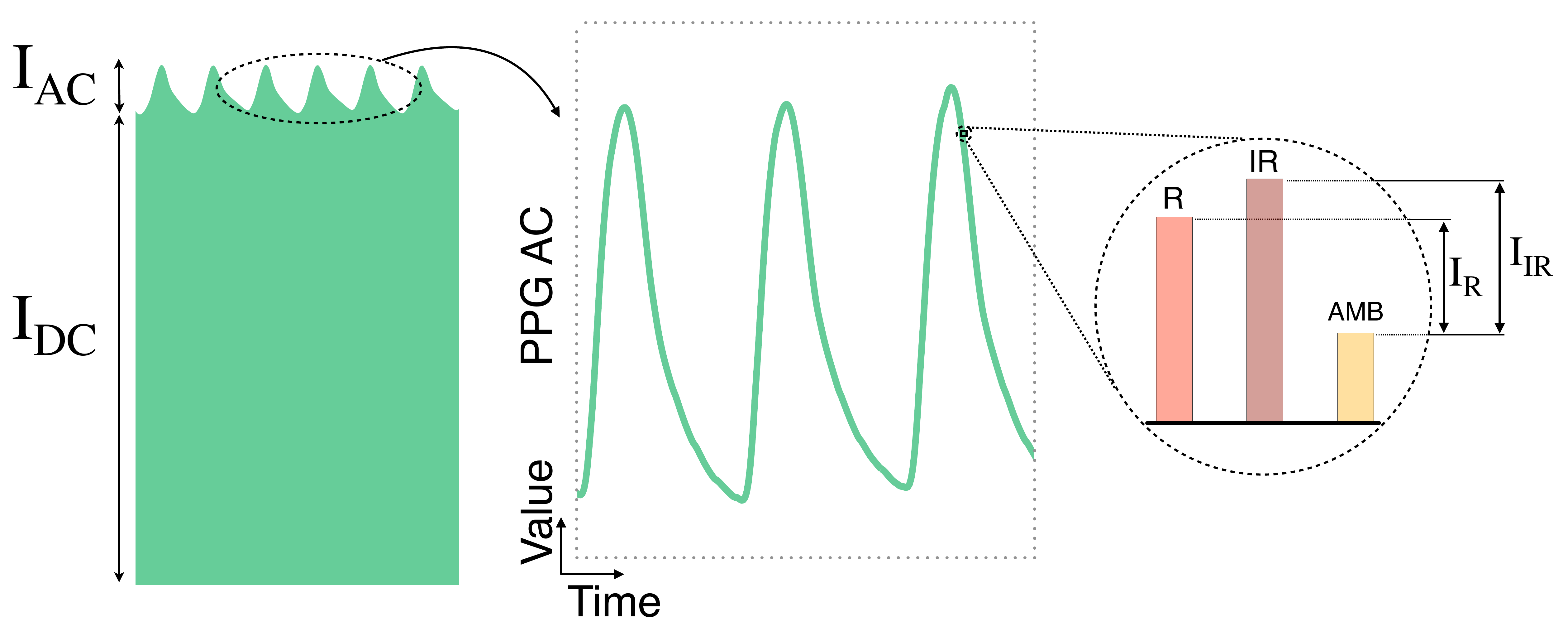}}
    \caption{PPG signal AC and DC components. Red and IR signals are computed after subtracting the AMB sample, performing system level CDS.}
    \label{fig2}
\end{figure}

\begin{figure}[!t]
    \setlength\abovecaptionskip{-0.3\baselineskip}
    \centering{\includegraphics[width=\columnwidth]{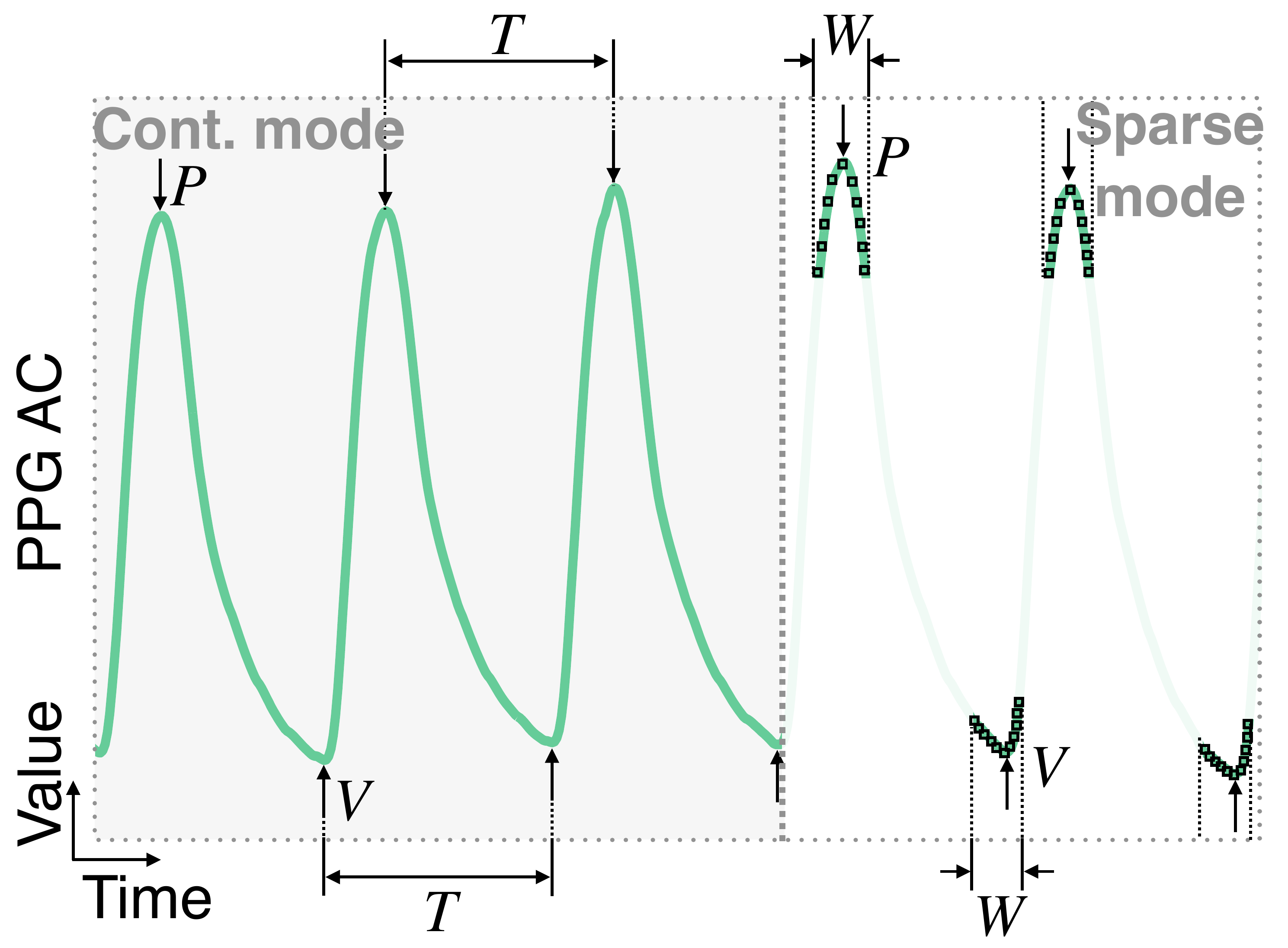}}
    \caption{Sparse sampling Algorithm. The sensor transitions to sparse mode after learning T over multiple cycles where it predicts next PAVs.}
    \label{fig3}
    \end{figure}

Large values of C$\mathrm{_{Par}}$ are an obstacle in achieving low input referred noise in current sensing frontends. Furthermore, C$\mathrm{_{Par}}$ poses a constraint on the achievable transimpedance gain and bandwidth in the TIA. As a result, handling very large values of C$\mathrm{_{Par}}$ as in OPDs necessitates careful design of the TIA and readout chain. Prior arts have used both capacitive and resistive feedback types in the TIA architectures. Traditionally, TIAs with capacitive feedback (CTIA) are deployed to achieve the lowest input referred current noise \cite{b16}. However, most CTIA designs only handle a few pFs of parasitic capacitance at the TIA input. In order to acquire the current signal out of OPDs with up to 10 nF of C$\mathrm{_{Par}}$, such topologies would require prohibitively large values of feedback and load capacitors ($>$100 pF) to deliver the required noise performance while realizing a reasonable gain. This can result in very large area occupation and demand larger transconductances, increasing the overall power consumption of the TIA.

In Resistive-Capacitive TIAs (ZTIA) however, the presence of feedback resistor enhances the feedback factor at lower frequencies. This causes the output noise spectrum to only increase at high frequencies, a phenomenon known as “noise peaking”. The output noise PSD of the ZTIA can be written in a simplified form as follows:
\begin{equation}
    \begin{gathered} S_{N_{out}}(f) \approx 4kT \left( \frac{1}{g_m}+R_F \right)
    \\
    \times \left( \frac{4\pi^2 \frac{R_F}{g_m} C_{Par}^2 f^2 + 1}{ 16\pi^4  C_F^2 C_{Par}^2 \frac{R_F^2}{g_m^2} f^4 + 4\pi^2 \left( \frac{C_{Par}^2}{g_m^2} + R_F^2 C_F^2 \right)f^2 + 1} \right)
    \end{gathered}
    \label{eq1}
\end{equation}
with $g_m$, $R_F$, and $C_F$ representing the effective transconductance of the OTA, the feedback resistance and capacitance. Only the OTA thermal noise as well as the feedback resistor’s Johnson noise is considered in this analysis. The bandwidth of this noise peaking can be greater than the system’s observation bandwidth. It is therefore possible to attenuate the TIA’s high frequency noise by means of sufficient filtering through the subsequent blocks. The appendix section covers the derivation of (1) as well as the comparison between CTIA and ZTIA topologies in greater detail. As a result, a ZTIA readout chain can potentially deliver a lower input referred noise compared to CTIA based architectures when C$\mathrm{_{Par}}$ is very large and is hence chosen in this design.

Like OPDs, OLEDs tend to have very large parasitic capacitances that take time to charge up, increasing the system settling time. Therefore, these parasitic elements slow down the sensor and limit the minimum duty cycle ratio of the system, thereby demanding more power. In addition, OLEDs typically require higher drive voltages up to 8 V \cite{b9} that further increase the sensor power consumption. It is therefore crucial to come up with strategies to alleviate the power requirements. The following section discusses a specific sparse sampling algorithm, implemented in this work, that aims to reduce the number of samples in the system and consequently save power.

\section{Sparse Sampling of PPG Signal}
\label{sec:sparse}
\subsection{HR and SpO$_2$ Extraction}

Figure 2 shows a typical PPG waveform containing both AC and DC components. The period (T) of the AC component represents a heart-beat cycle and can be used to compute HR. Thus, PPG using a single LED is sufficient to provide HR information. Prior arts have routinely used the average value of the period over 2 or 8 second windows. In this paper, an 8 second window is selected to report HR. (Eq. 2)
\begin{equation}
    HR=\frac{60}{\overline{T}}
\end{equation}

SpO$_2$, however, depends on relative concentration of oxygenated versus de-oxygenated hemoglobin, and hence, needs information at two different light wavelengths. As light passes through the tissue, it is attenuated by different elements and this attenuation is described by the extinction coefficient. Light at Red and IR wavelengths have been used traditionally \cite{b24} since they offer the largest difference in extinction coefficients when passing through oxygenated and de-oxygenated hemoglobin. The value 
\begin{figure*}
    \setlength\abovecaptionskip{-0.3\baselineskip}
    \includegraphics[width=\textwidth]{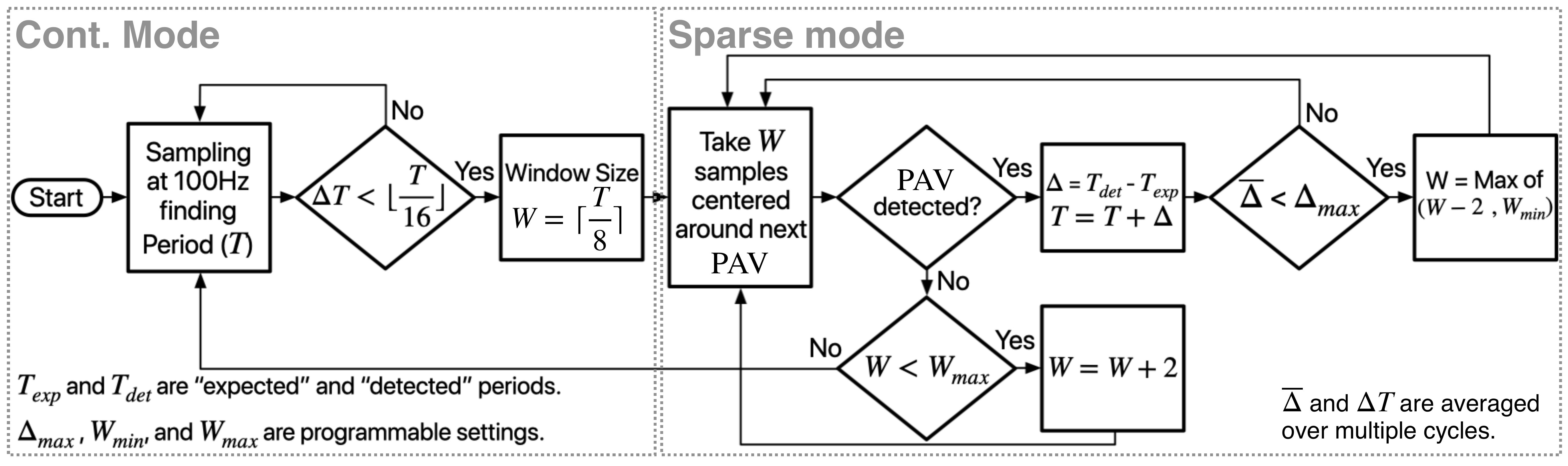}
    \caption{Sparse sampling algorithm flow-chart.}
    \label{fig4}
\end{figure*}
of SpO$_2$ is computed based on the ratio of AC component over the DC component of Red and IR PPG waveforms (Eq. 3-4). \cite{b9}
\begin{equation}
    R_{OS}=\frac{I_{AC_{Red}}}{I_{DC_{Red}}} / \frac{I_{AC_{IR}}}{I_{DC_{IR}}}
\end{equation}
\begin{equation}
    SpO_2 = \frac{\varepsilon_{Hb_{Red}} - \varepsilon_{Hb_{IR}}\cdot R_{OS}}{\varepsilon_{Hb_{Red}} - \varepsilon_{HbO2_{Red}} + [\varepsilon_{Hb_{IR}} - \varepsilon_{HbO2_{IR}}]\cdot R_{OS}} 
\end{equation}
where $\varepsilon$ is the extinction coefficient of light at Red or IR wavelength through oxygenated or de-oxygenated hemoglobin. To compute SpO$_2$, only the peak and valley (PAV) values of the PPG signals are needed in the two wavelengths. It is therefore possible to only sample the signal PAVs and as a result save power.
\subsection{Sparse Sampling Algorithm}

The idea of the proposed sparse sampling algorithm is shown in Figure 3 where the sensor has two modes of operation. The flowchart of this sparse sampling algorithm is also shown in Figure 4. The sensor starts by uniformly sampling the PPG signal at 100Hz (continuous mode). The period of the AC component (T) is learned over multiple cycles. Once a stable value is realized, the sensor enters the sparse mode where it predicts the upcoming PAVs and takes a few samples centered around them. A window size (W) of $\lceil$T/8$\rceil$ is initially selected which can then be reduced upon successful detection of PAVs. A smaller W means that fewer samples are taken and as a result the overall power is decreased. However, larger W allows the backend to accurately detect PAVs in the presence of high heart rate variability.
Therefore, cycle-to-cycle updates are made to the estimated value of T to adjust for any period drift. If PAVs are repeatedly missed due to rapid drifts or large motion artifacts, the sensor expands its observation window by increasing W until a programmable maximum, W$_Max$. Without new PAVs detected, the sensor can revert to the continuous mode to re-learn T. The computed T, together with samples taken from signal’s PAVs are adequate to provide both HR and SpO$_2$ information. 
\begin{figure}[!b]
    \setlength\abovecaptionskip{-0.3\baselineskip}
    \centering{\includegraphics[width=\columnwidth]{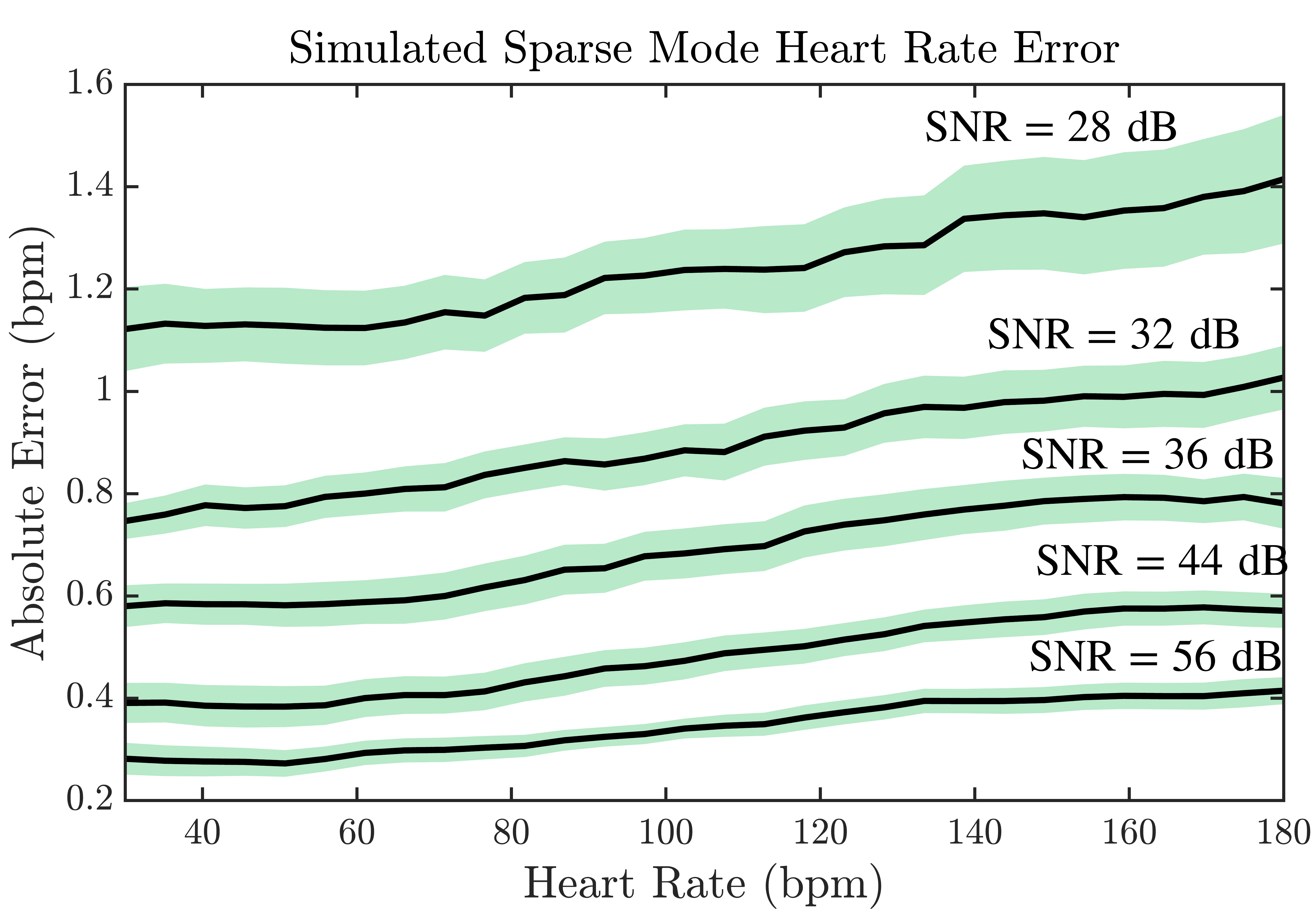}}
    \caption{Simulated sparse mode HR error for input sine waves at different frequencies and SNRs. Green shades show the $\pm 3\sigma$ range of the error.}
    \label{fig5}
\end{figure}

Proper detection of PAVs by the backend depends on the SNR of the sparse samples as well as how fast the input is changing. To evaluate the performance of the proposed algorithm at different SNR levels and a range of HR values, a set of MATLAB simulations are run that quantify the reported HR error. The results of such simulations provide information about the required SNR levels at different heart rates to achieve the desired 1 bpm accuracy. These simulations were run with a sine wave input at frequencies ranging from 0.5 to 3 Hz, corresponding to HRs of 30 bpm to 180 bpm, and at amplitudes corresponding to SNRs from 28 dB to 56 dB. In each run, sparse sampling is performed on the sine wave input and the estimated periods are recorded, averaged over 8 second windows. A total of 12,800 simulations were conducted to capture the mean and variance of the reported HR errors. The results are shown in Figure 5, where the solid lines show the mean error, and the green shades represent the $\pm3\sigma$ of the computed HR errors. The results confirm that the proposed algorithm provides a robust operation under a wide range of SNR and heart rates. Furthermore, the results demonstrate that a minimum SNR of $\sim$30 dB is sufficient in achieving sub-1 bpm HR error for HRs up to 180 bpm.  A similar test is performed in measurements to quantify the performance of the sensor and the results will be discussed in section V.

\section{Circuit Implementation}
\subsection{Frontend architecture}

Figure 6(a) shows the detailed block diagram of the transimpedance frontend (TFE) as well as the on-chip digital backend (DBE). The PD is connected and read out differentially at the input. This eliminates the need for a separate low noise bias voltage on the other end of the detector. The 0 V bias also minimizes the PD dark current. 8-bit (5-bit thermometer/3-bit binary coded) differential I-DACs (P and N-DACs) subtract the DC component at the input up to 15 $\mu$A. The I-DAC LSB reference current is tunable from 20 nA to 60 nA to provide a wider range and granularity for the subtraction. N and P I-DAC codes are computed separately to adjust for any mismatch and gain error between them. This calibration only occurs once and the resulting coefficients are programmed into the backend. The remaining AC component of the signal is then amplified via a differential ZTIA followed by a reset integrator that provides additional gain and boxcar averaging, obviating the need for an explicit anti-alias filter. A 12-bit synchronous SAR ADC samples the integrator output and delivers the digital code to the DBE.
\begin{figure}[!t]
    \setlength\abovecaptionskip{-0.3\baselineskip}
    \centering{\includegraphics[width=\columnwidth]{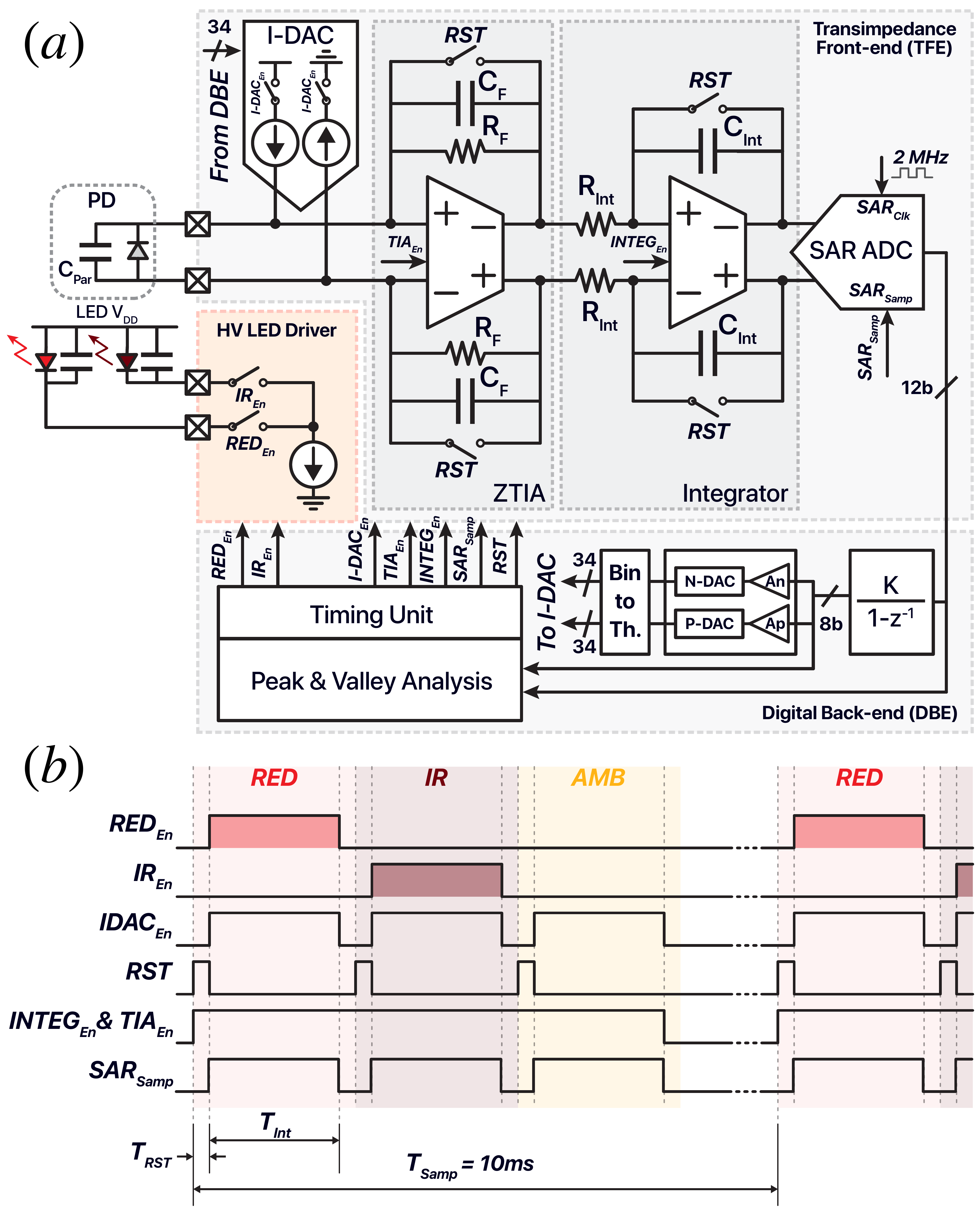}}
    \caption{Detailed channel block diagram (a) and timing diagram (b).}
    \label{fig6}
\end{figure}

The timing of TFE operation is diagrammed in Figure 6(b). Every sample consists of three back-to-back phases. Red and IR signals are acquired followed by an ambient (AMB) phase where no LED is on and the photocurrent due to the ambient light as well as the PD dark current are sampled. AMB must be sampled with each Red/IR sample since there can exist time-varying changes in the ambient lighting as well as reflected pulsatile components during this phase. AMB subtraction also serves as a system level correlated double sampling (CDS) that helps remove any offset and flicker noise \cite{b15}. The DC DAC code for each phase is computed, stored, and updated separately for fastest operation. T$\mathrm{_{RST}}$ and T$\mathrm{_{Int}}$ are programmable settings that set the reset and integration durations for TFE. In most measurements, T$\mathrm{_{Int}}$ is set from 25 $\mu$s to 100 $\mu$s with T$\mathrm{_{RST}}$ as short as 5 to 10 $\mu$s. An on-chip timing unit provides all timing signals to the TFE and LED driver sub-blocks to maintain synchronicity. A peak and valley analysis block within the DBE performs algorithm computations and triggers counters that determine the upcoming sampling windows.

\subsection{ZTIA Design}

As previously discussed, a ZTIA topology is utilized to accommodate a wide range of photodiode capacitance (C$\mathrm{_{Par}}$) while maintaining a high SNR. The use of a resistor in the feedback eliminates the reset noise (kT/C) and the need for a separate CDS phase. Figure 7 shows the schematic of the 3-stage current reuse core OTA. The current reuse topology
\begin{figure}[!t]
    \setlength\abovecaptionskip{-0.3\baselineskip}
    \centering{\includegraphics[width=\columnwidth]{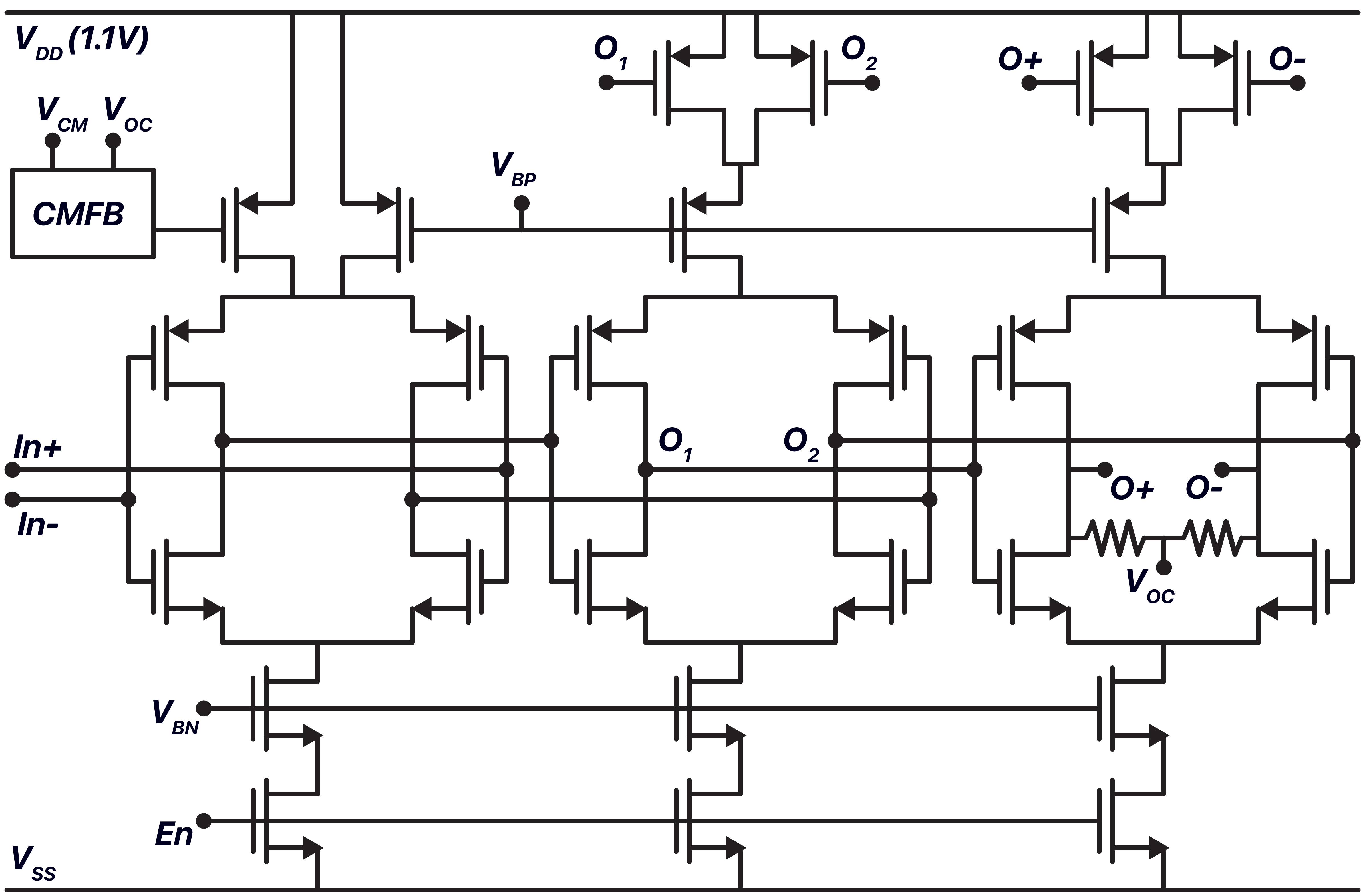}}
    \caption{Schematic of the ZTIA core OTA. Current reuse topology enhances current efficiency and reduces power.}
    \label{fig7}
\end{figure}
enhances the performance by doubling the effective current efficiency factor (g$\mathrm{_m}$/I$\mathrm{_D}$) of each stage. The fully differential architecture also provides better immunity against supply and common mode noise sources, as well as improved linearity. Local and global common mode feedback networks are included to provide common mode bias and improve common mode rejection. Reset switches connect the inputs and outputs of the TIA during the RST phases to quickly bring the outputs to the mid-rail bias point. Values of RF and CF are tunable to provide a gain of 1 to 5 M$\Omega$ and an adjustable bandwidth. The entire OTA is power gated using low resistance NMOS transistors to save power outside of sampling windows. 

\subsection{Reset integrator and SAR ADC}

The RC integrator captures the TIA output over T$\mathrm{_{INT}}$ and provides differential outputs to the SAR ADC sampling capacitors. A SAR topology is selected to achieve a low power consumption while providing the required resolution. The value of C$\mathrm{_{Int}}$ is tunable to allow for adjustable gain. The integrator is reset at the beginning of every phase via the RST switches. This integration and reset provides a boxcar averaging transfer function detailed in Eq. (5).

\begin{equation}
    H_{Sinc}(f) = \frac{T_{Int}}{R_{Int}\cdot C_{Int}} \cdot Sinc(T_{Int}\cdot f)
\end{equation}

This low pass transfer function significantly attenuates the high frequency noise content of the TIA output. The feedback capacitors are sized such that the reset kT/C noise of the integrator is insignificant when referred to the channel input. Similar to the OTA, the integrator is shut down outside of sampling windows to save power. A 12-bit synchronous SAR ADC is implemented using monotonic switching scheme that only needs half the total capacitance compared to traditional switching methods. The SAR ADC uses a 2 MHz reference clock provided by the DBE to perform the conversions.

\subsection{Peripheral circuits}

An on-chip PTAT current reference is used to supply the bias current of all analog sub-blocks. The PTAT nature of the reference increases the bias current in proportion to the temperature to maintain a constant noise level in the readout chain. A high voltage (HV) LED driver
\begin{figure}[!t]
    \setlength\abovecaptionskip{-0.3\baselineskip}
    \centering{\includegraphics[width=\columnwidth]{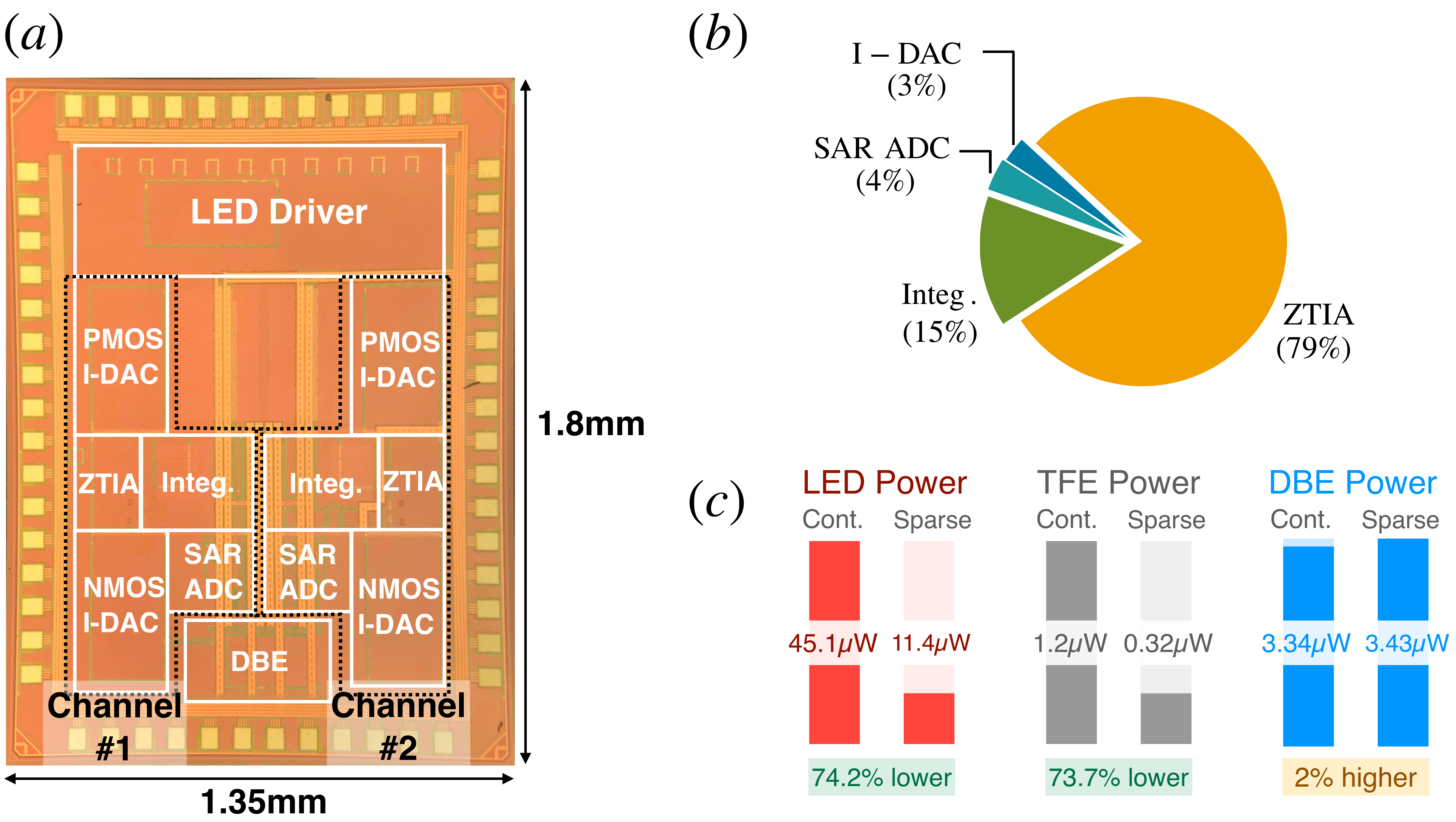}}
    \caption{Chip micrograph (a). TFE power breakdown in continuous mode. (b) System power reduction between continuous mode and sparse mode. (c)}
    \label{fig8}
\end{figure}
is implemented on-chip to drive the LEDs with up to 16 mA of current. The driver uses HV transistors that can operate with up to 8 V of supply voltage, enabling the use of OLEDs. The driver, which is controlled by the timing unit, sinks the current from the LEDs. The drive strength is 5-bit tunable and is separately set for Red and IR phases.

\section{Measurement Results}

\subsection{Chip Micrograph and Power Breakdown}

The IC was fabricated in a TSMC 40 nm HV technology and occupies an area of 1.35 $\times$ 1.8 mm$^2$. Figure 8 shows the chip micrograph as well as the TFE and system power breakdown. In continuous mode, the sensor consumes a total of 49.7 $\mu$W with 45.1 $\mu$W, 1.22 $\mu$W, and 3.34 $\mu$W drawn by the two LEDs, the TFE, and the DBE respectively. Sparse mode lowers the system power consumption to 15.2 µW with LEDs and TFE power going down by $\sim$75\%, and the DBE power increasing by only 2\%. 
\begin{figure}[!t]
    \setlength\abovecaptionskip{-0.3\baselineskip}
    \centering{\includegraphics[width=\columnwidth]{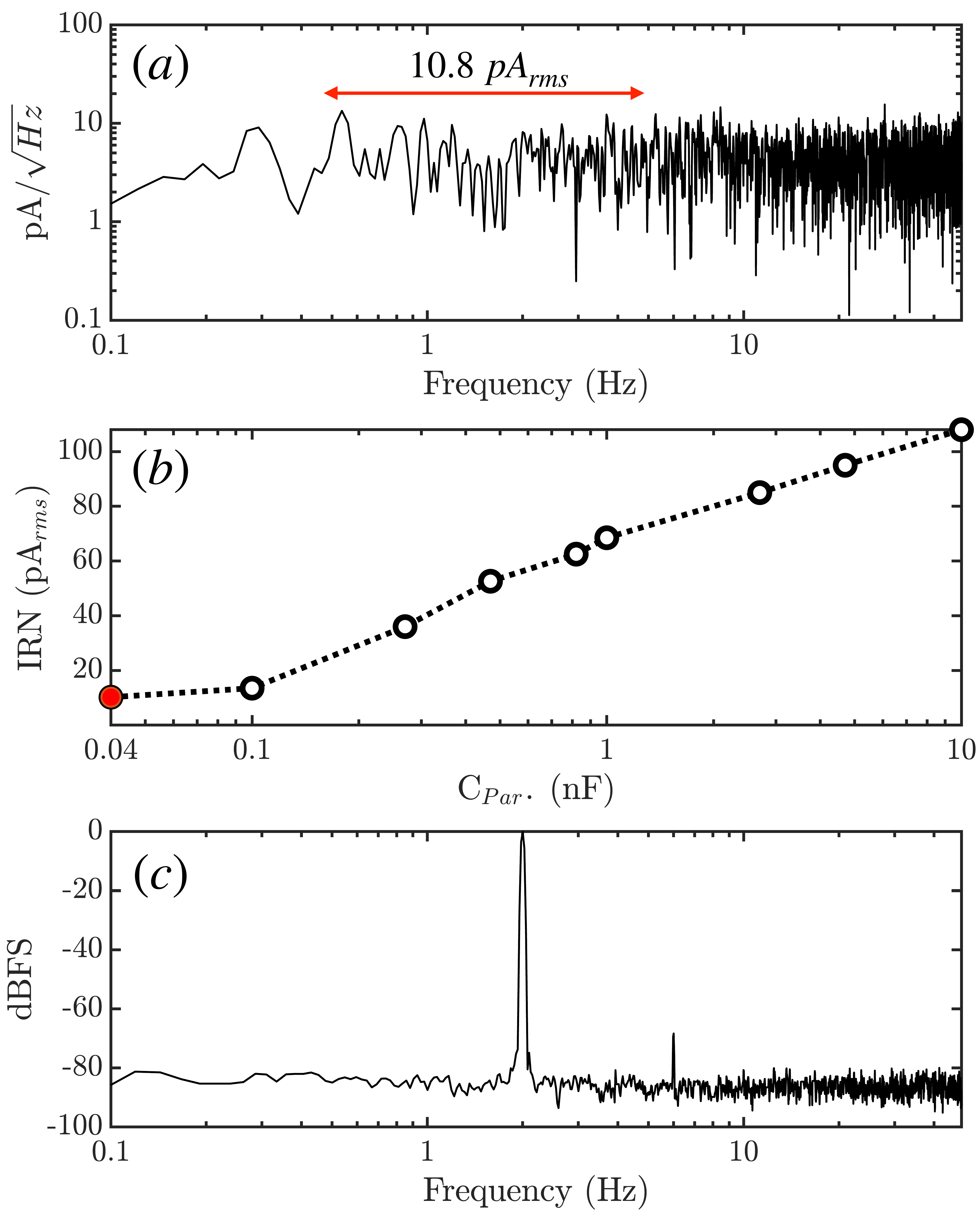}}
    \caption{Electrical testing results. (a) IRN spectrum for a 40 pF C$\mathrm{_{Par}}$. (b) Integrated IRN over 5 Hz bandwidth vs.  C$\mathrm{_{Par}}$. The red dot corresponds to the spectrum from (a) (c) ADC output spectrum for a 50 nA$_\mathrm{pp}$ 2 Hz sine wave input.}
    \label{fig9}
\end{figure}
The pie chart in Figure 8 shows the breakdown of power consumption of the individual sub-blocks within the TFE in continuous mode where most of the power is dissipated in the ZTIA to maintain a low input referred noise.

\subsection{Electrical Measurements}

The performance of the TFE was characterized via benchtop electrical measurements and the results are presented. Figure 9(a) shows the input referred noise (IRN) spectrum of the readout at 40 M$\Omega$ of overall gain. The measurement was performed with a C$\mathrm{_{Par}}$ = 40 pF matching the capacitance of the commercial PD. It achieved a noise spectrum density of 4.8 pA/rtHz and 10.8 pA$_\mathrm{rms}$ integrated noise over the 5 Hz bandwidth. IRN increased with C$\mathrm{_{Par}}$ as plotted in Figure 9(b) for C$\mathrm{_{Par}}$ as large as 10 nF, the largest measured capacitance of the OPDs \cite{b9}. The ADC output spectrum for a 2 Hz sine wave input with 50 nA$_\mathrm{pp}$ of amplitude is shown in Figure 9(c). The TFE achieved an SFDR of 68.3 dB and an SNDR of 62.4 dB with the 3$^{\mathrm{rd}}$ harmonic forming the largest spur.

The simulation results from section III (Figure 5) were verified via similar benchtop measurements. A set of 48 measurements were performed while providing a sine wave to the TFE with sparse mode enabled. The frequency of the sine wave was swept from 0.5 to 3 Hz, corresponding to 30 to 180 bpm HR. Two different amplitudes were selected for the input sine wave to achieve two distinct SNR levels, 32 dB and 44 dB. Figure 10 presents the measured effective HR error for every experiment. The measured results are well in agreement with the simulated numbers, confirming the performance and robustness of the algorithm. Less than 1 bpm of average error was measured with an SNR as low as 32 dB for HR up to 180 bpm.

\begin{figure}[!t]
    \setlength\abovecaptionskip{-0.3\baselineskip}
    \centering{\includegraphics[width=\columnwidth]{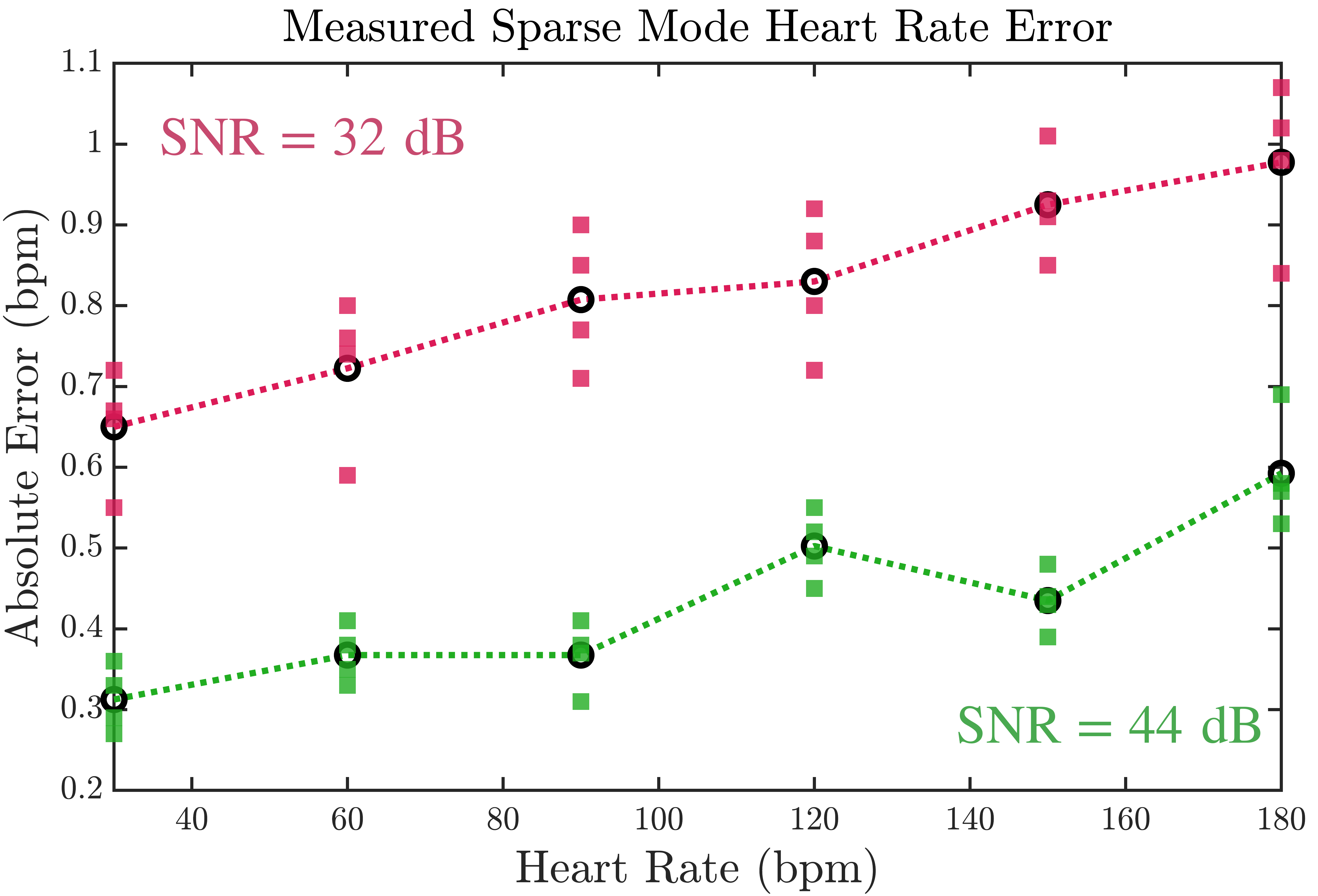}}
    \caption{Measured sparse mode HR error with input sine waves at different frequencies and at two levels of SNR. Black circles show the average error at each rate. The computed error is the mean absolute error over 8 s windows. }
    \label{fig10}
\end{figure}

\subsection{In vivo Results}

A set of \textit{in vivo} experiments were performed with the sensor in both continuous and sparse modes of operation. In these experiments, the sensor was placed on the index finger of a healthy adult in a sitting position, under typical incandescent lighting and at room temperature. To evaluate the accuracy of the sensor output, a clinical grade pulse-oximeter (Wellue HPO) was attached to the subject’s ring finger to perform measurements simultaneously with the sensor IC. In the first 
\begin{figure}[!t]
    \setlength\abovecaptionskip{-0.3\baselineskip}
    \centering{\includegraphics[width=\columnwidth]{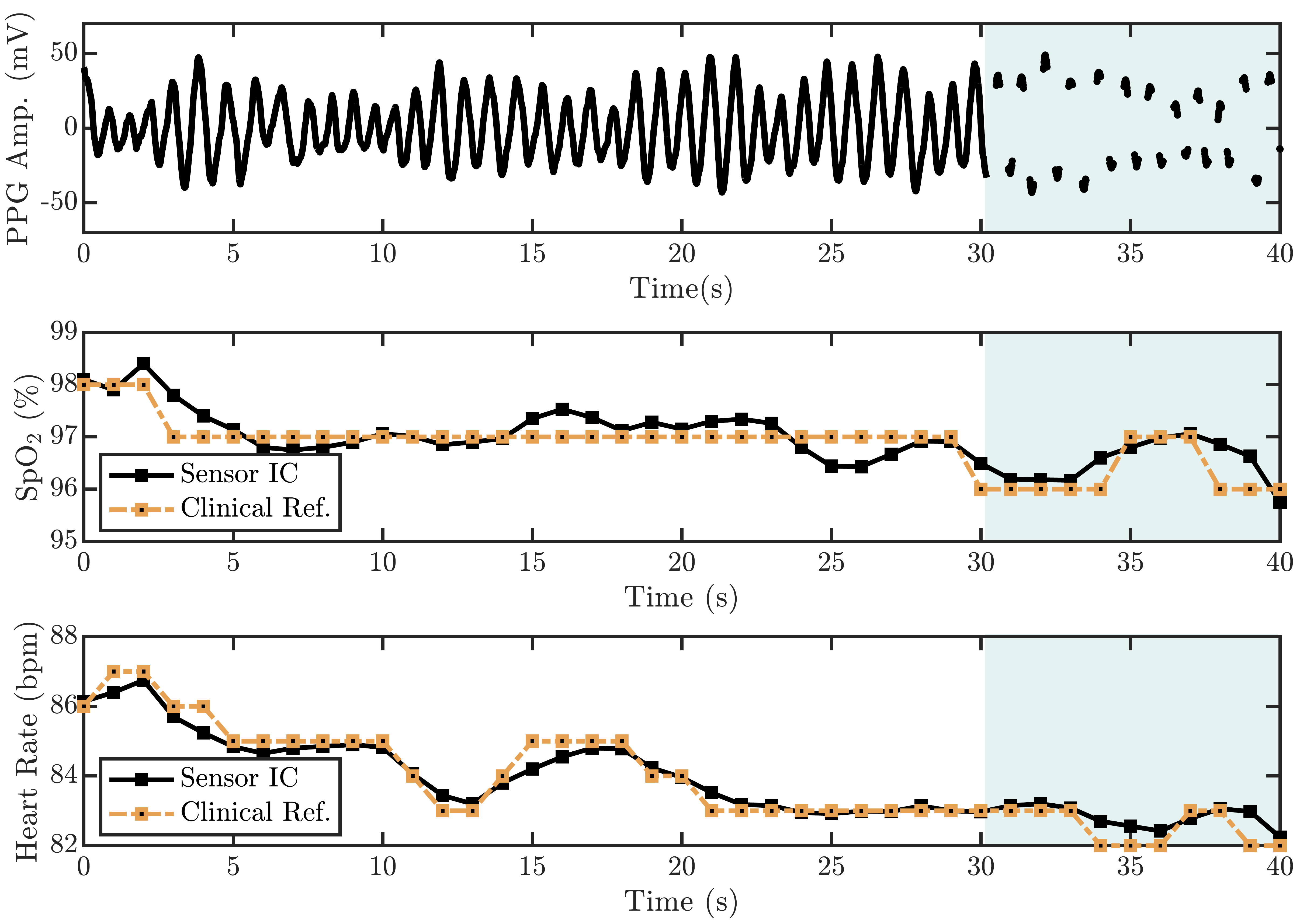}}
    \caption{\textit{In vivo} verification of the sensor IC in both continuous and sparse modes using commercial silicon PD and LEDs. The PPG waveform and its corresponding HR and SpO$_2$ measurements are shown.}
    \label{fig11}
\end{figure}
experiment, commercial PDs and LEDs were used as interface devices and the PPG signal, the HR, and the SpO$_2$ results were recorded. Figure 11 plots these results against data from the clinical reference. The accuracy of the sensor was maintained after transitioning to sparse mode where the HR and SpO$_2$ mean absolute errors only rose from 0.3 bpm and 0.5\% to 0.4 bpm and 0.7\% respectively.

\begin{figure}[!b]
    \setlength\abovecaptionskip{-0.3\baselineskip}
    \centering{\includegraphics[width=\columnwidth]{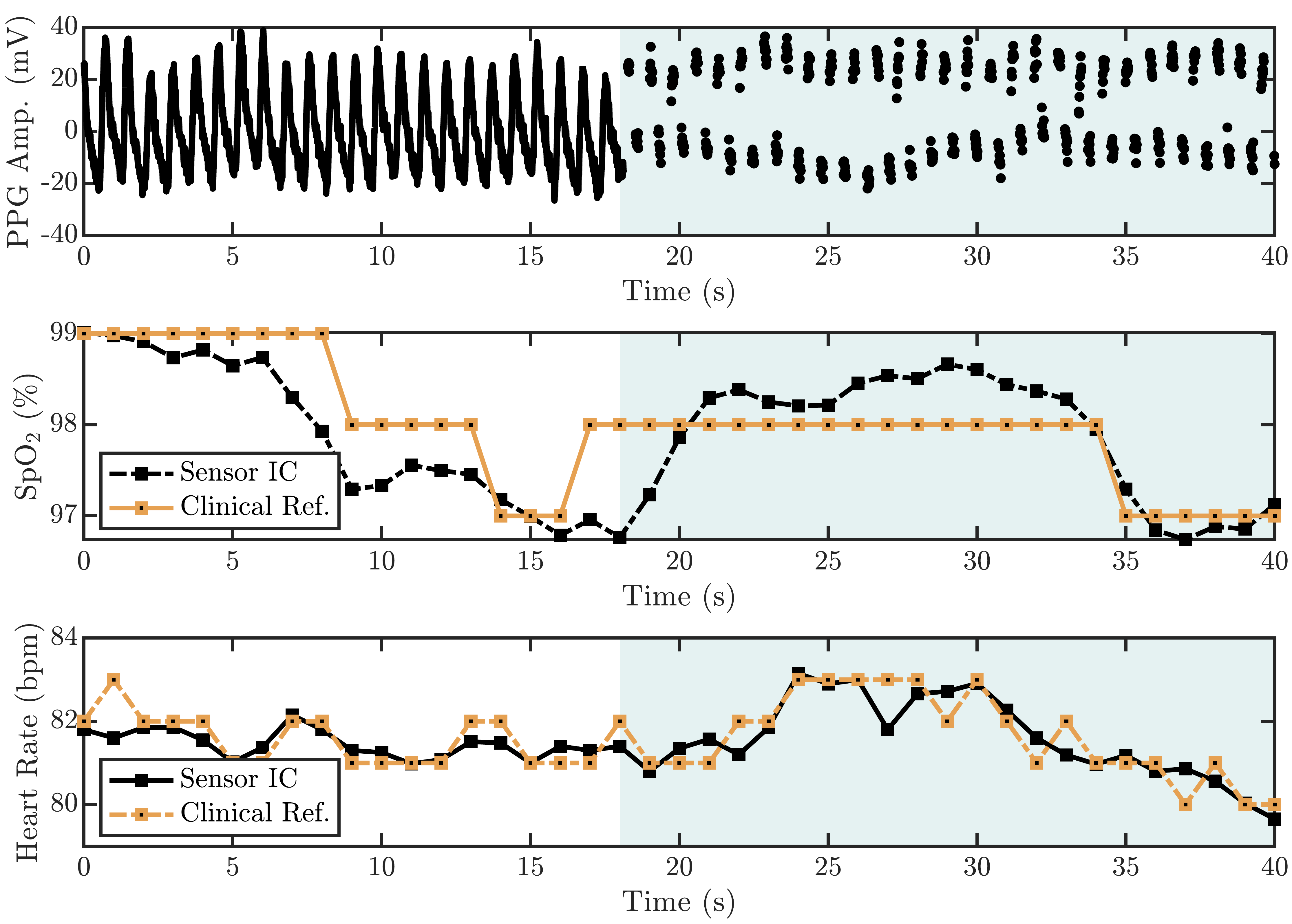}}
    \caption{\textit{In vivo} verification of the sensor IC in both continuous and sparse modes using Organic Interface Devices (OLEDs \& OPDs). The PPG waveform and the corresponding HR and SpO$_2$ measurements are shown.}
    \label{fig12}
\end{figure}

A similar experiment was performed using a set of flexible organic devices discussed in [23]. The PPG waveform, and the corresponding measured HR and SpO$_2$ are plotted in Figure 12 in both continuous and sparse modes. The sensor achieved less than 1 bpm and less than 1\% HR and SpO$_2$ errors when compared against the clinical reference.
\begin{figure}[!t]
    \setlength\abovecaptionskip{-0.3\baselineskip}
    \centering{\includegraphics[width=\columnwidth]{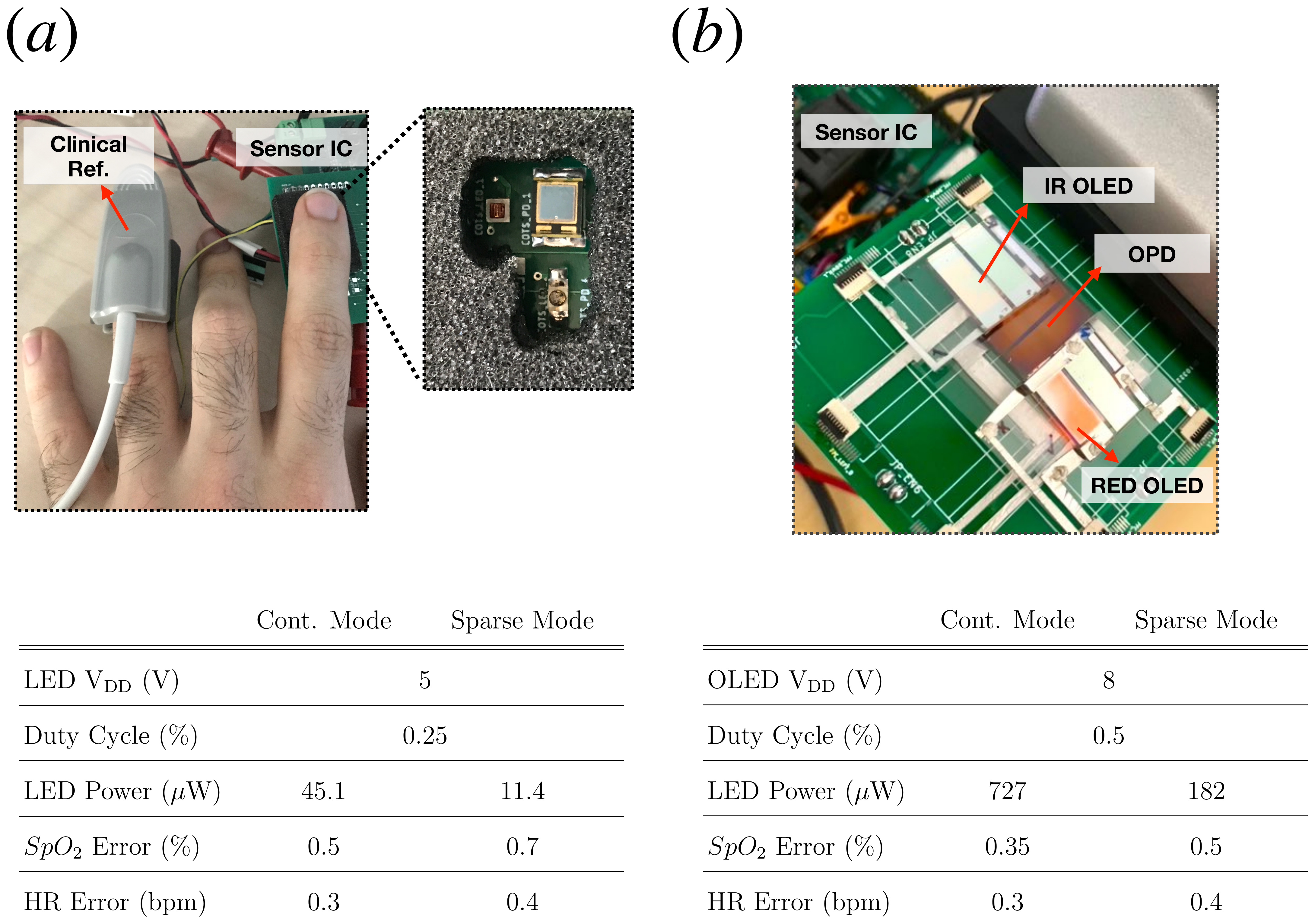}}
    \caption{\textit{In vivo} measurement setups as well as a summary of the results when using (a) commercial silicon devices and (b) flexible organic devices. }
    \label{fig13}
\end{figure}
The use of organic devices required a higher drive voltage (8 V), an increased drive current as well as a higher (0.5\%) duty cycle ratio resulting in overall higher power consumption. However, enabling sparse mode significantly lowered the OLED power by about 75\%, improving sensor battery life. Figure 13 presents a summary of the \textit{in vivo} results together with the measurement setups when using (a) commercial and (b) organic devices.

\begin{figure}[!t]
    \setlength\abovecaptionskip{-0.3\baselineskip}
    \centering{\includegraphics[width=\columnwidth]{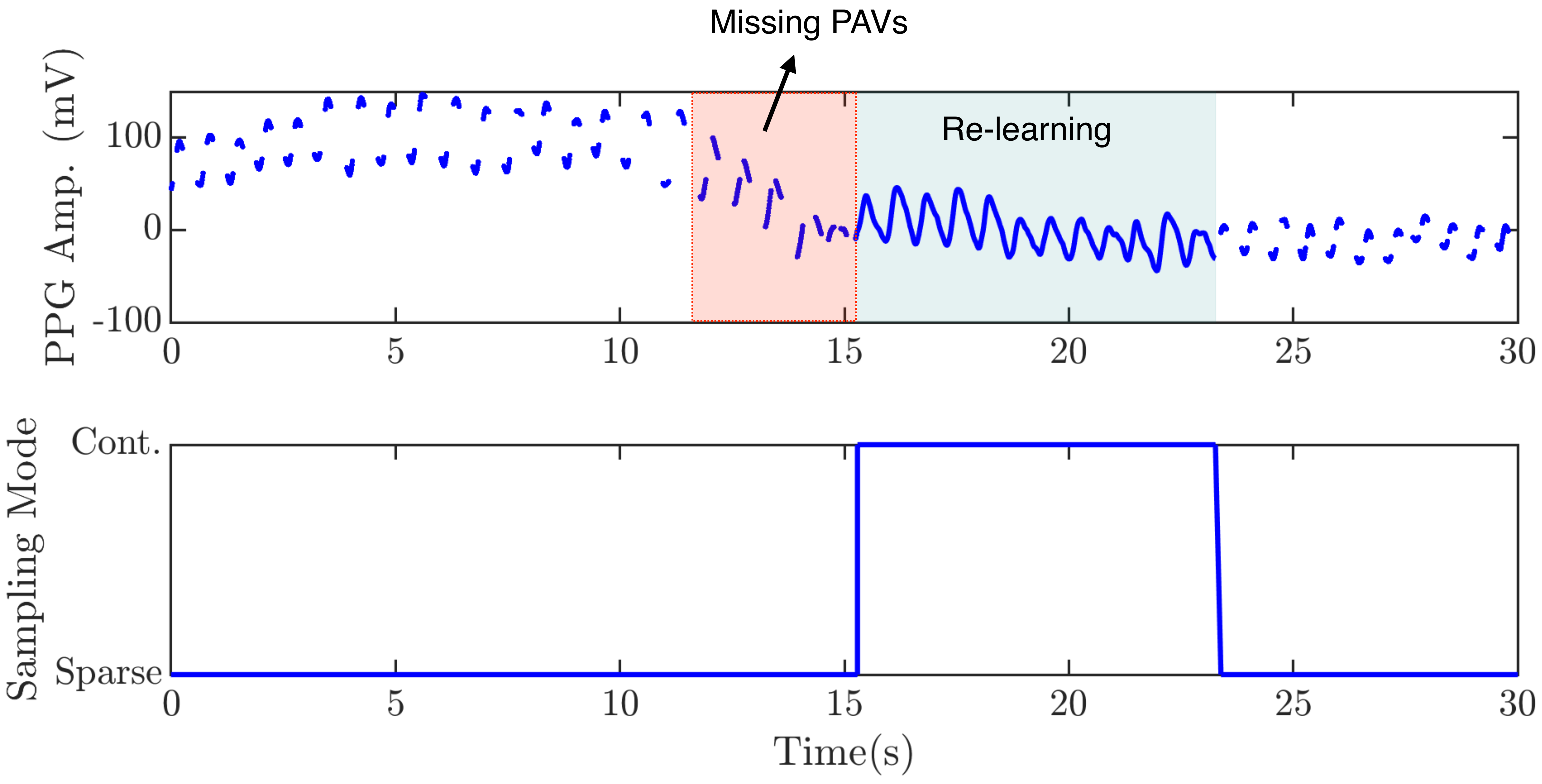}}
    \caption{Missing PAVs due to motion artifact creating a large change in the sampled PAV values. The backend initially expands W, then reverts to the continuous mode and finally re-gains lock on the PPG.}
    \label{fig14}
\end{figure}
\renewcommand{\arraystretch}{1.5} 
\begin{table*}
\caption{Comparison with prior arts of PPG and SpO$_2$ sensors}
\centering
\begin{tabular}{l|c|c|c|c|c|c|c|c|cc} 
    \hline
    \hline
 Reference & ISSCC & TBCAS & TBCAS & TBCAS & TBCAS & ISSCC & TBCAS & JSSC & \multicolumn{2}{c}{This work}   \\ 
 \cline{10-11} 
        & 2016 & 2017 & 2018 & 2019 & 2019 & 2021 & 2021 & 2021 & Continuous   & Sparse \\
        & [8] &  [19] &  [20] & [26] & [14] & [17] & [25] & [18] & Mode   & Mode
        
        \\ \hline
Technology (nm)   &  180 & 180 & 180 & 55 & 180 & 65 & 180 & 65 & \multicolumn{2}{c}{40 HV}  \\    
    \hline
 
 V$_{\mathrm{DD}}$ (V) [LED/Readout]   & 5/1.5 & 5/1.2 & 3.3 & 2.8/1.2 & 3.3/1.8 & 1.8/1 & 3.3/1.2 &  -/0.6 & \multicolumn{2}{c}{5/1.1} \\ 
    \hline
 
 \multirow{2}{12em}{Readout Power ($\mu$W/Ch.)}    &  \multirow{2}{*}{87$^{a}$} & \multirow{2}{*}{172} & \multirow{2}{*}{27.4} & \multirow{2}{*}{54} & \multirow{2}{*}{9} & \multirow{2}{*}{24} & \multirow{2}{*}{28} & \multirow{2}{*}{0.532} &   TFE: 1.22 & TFE:0.32 \\
 & & & & & & & & & DBE: 3.34 & DBE: 3.43 \\
 \hline
 
 LED Power ($\mu$W/LED)   &   $~ ~ ~$  27$^{a,d}$    & $~$120$^{c}$ & 16 & 102.5 & 17.5 & 11.5 & 305 & 8.5 & 22.6 & 5.7 \\ 
 \hline
 Total Power ($\mu$W/Ch.)   & 114 & 292 & 43.4 & 156.5 & 26.5 & $~$35.5$^{a}$ & 333 & 9.032 & 27.2 & 9.45 \\ 
 \hline
 
 Sampling Freq. (Hz)   & 400 & 128 & 100 & 128 & 25 & 20 & 2048 & 20 & \multicolumn{2}{c}{100}  \\ 
 \hline
 
 Duty Cycle (\%)   & 2 & 0.4 & 0.25 & - & 0.7 & 0.04 & 1 & 1 & \multicolumn{2}{c}{0.25}  \\ 
 \hline
 
 Input Noise (pA/$\sqrt{Hz}$)   & - & 153 & - & 6.3 & - & - & $~$8.7$^{a}$ & 20.1 & \multicolumn{2}{c}{$~$4.8$^{e}$}  \\ 
 \hline
 $SpO_2$ Error (\%)   & $~$1.1$^{b}$ & - & - & - & - & - & - & - & 0.5$^{f}$ & 0.7$^{f}$  \\ 
 \hline
 
HR Error (bpm)   & - & $~$2$^{b}$ & $~$2$^{b}$ & -  & $~$1.4$^{f}$ & $~$1.9$^{f}$ & - & - & 0.3$^{f}$ & 0.4$^{f}$ \\ 
\hline
\hline
\end{tabular}
\begin{tabbing}
    {$^{{a}}$Estimated} \hspace{1pt}  
    {$^{{b}}$Max. error} \hspace{1pt}
    {$^{{c}}$ $10\times$ Compression Rate} \hspace{1pt} 
    {$^{{d}}$Organic LEDs/PDs} \hspace{1pt} 
    {$^{{e}}$ $C_{par}$ = 40pF (Matching $C_{pd}$)} \hspace{1pt}
    {$^{{f}}$Mean Abs. error.} \hspace{1pt} \\
\end{tabbing}
\vspace{-5mm}
\end{table*}
Figure 11 captures an incident where the sparse mode operation is interrupted by a motion artifact that is large enough to create a big difference in sampled PAV values compared to the previously captured PAVs. This causes an unwanted shift in the estimated period and results in the following PAVs to be missed. As shown in the figure, the sensor initially tries to find new PAVs by expanding the observation window and increasing W. Since no new PAVs are found after W reaches W$_\mathrm{Max}$, the sensor eventually reverts to continuous mode. Shortly after this transition occurs, the backend re-learns the signal period and the system re-enters sparse mode.

\section{Summary}

A comparison against the most recent prior arts of PPG and SpO$_2$ sensors is provided in Table I. The LED power is normalized by the number of LEDs in the system since only a single wavelength PPG measurement is required to report HR, while SpO$_2$ requires two. Compared to state-of-the-art, this work achieves the lowest LED power and one of the lowest total power consumptions while simultaneously delivering the lowest input referred noise. The measured HR and SpO$_2$ errors are less than 1 bpm and 1\% respectively and are lower compared to other works.

The functionality and performance of the proposed sensor IC was characterized using both commercial and flexible organic interface devices. The \textit{in vivo} measurement results confirm the accuracy of the sensor data when operating with a wide range of optical components including organic devices with parasitic capacitances as large as 10 nF. The introduction of reconstruction-free sparse sampling reduces the overall system power by nearly 70\% while maintaining the accuracy of the output data.
\appendix[Noise Analysis, ZTIA vs. CTIA]


This section covers the analytical description of the TIA noise when operating with a large C$\mathrm{_{Par}}$ at the input. For simplicity, the core OTA is modeled as a single stage with a transconductance of g$\mathrm{_m}$. Figure 15 shows the representative diagram of the discussed circuit. The two major sources of noise are the amplifier thermal noise modeled as a current source at the OTA output as well as the thermal noise of R$_\mathrm{F}$. Each of these sources impact the TIA output voltage with their specific transfer functions, shown in simplified forms as follows (Eq. 6 \& 7).
\begin{equation}
    H_{v_{n_{R_F}}}(s) = \frac{C_{Par} s + g_m}{C_{Par}C_F R_F s^2 + (C_{Par} + C_F g_m R_F) s + g_m}
\end{equation}
\begin{equation}
    H_{i_{n_{g_m}}}(s) = \frac{R_F (C_{Par} + C_F) s + 1}{C_{Par}C_F R_F s^2 + (C_{Par} + C_F g_m R_F) s + g_m}
\end{equation}
The power spectral densities (PSD) of the two noise sources can be written as:
\begin{equation}
    \frac{v_{n_{R_F}}^2}{\Delta f} = 4kT R_F, \qquad \frac{i_{n_{g_m}}^2}{\Delta f} = 4kT\gamma \alpha g_m
\end{equation}
Where $\alpha$ and $\gamma$ are parameters relating to the technology and topology of choice and $T$ is the absolute temperature.
\begin{figure}[!t]
    \setlength\abovecaptionskip{-0.3\baselineskip}
    \centerline{\includegraphics[width=\columnwidth]{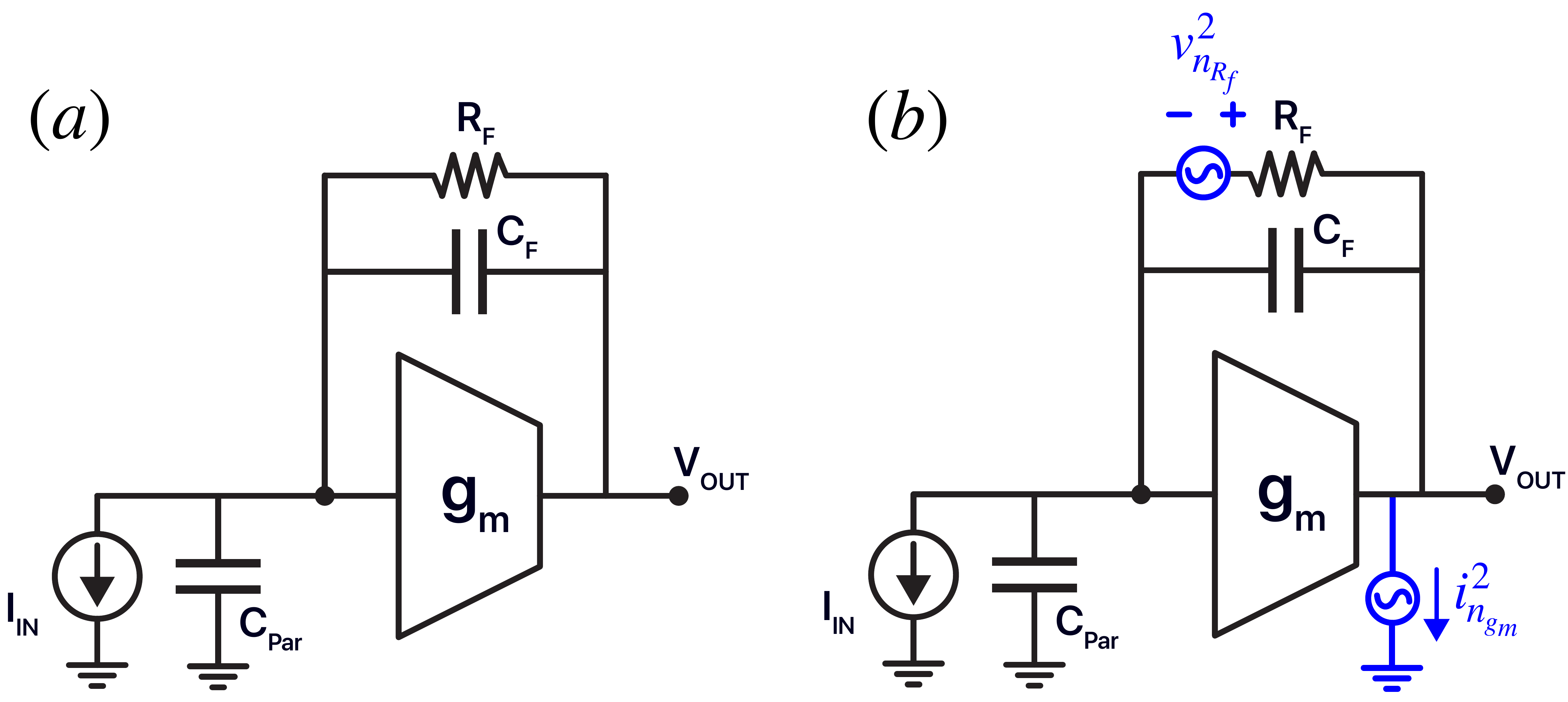}}
    \caption{Simplified diagram of the ZTIA used for the analysis (a). The noise sources of the OTA and RF are shown in (b).}
    \label{fig15}
\end{figure}
For simplicity, we assume $\alpha = 1$, $\gamma = 1$. This gives the following PSD for the TIA output noise:
\begin{equation}
    \begin{gathered} S_{N_{out}}(f) \approx 4kT \left( \frac{1}{g_m}+R_F \right)
    \\
    \times \left( \frac{4\pi^2 \frac{R_F}{g_m} C_{Par}^2 f^2 + 1}{ 16\pi^4  C_F^2 C_{Par}^2 \frac{R_F^2}{g_m^2} f^4 + 4\pi^2 \left( \frac{C_{Par}^2}{g_m^2} + R_F^2 C_F^2 \right)f^2 + 1} \right)
    \end{gathered}
    \label{eq2}
\end{equation}
At low frequencies, the PSD has a value of,
\begin{equation}
    S_{N_{out}}(f) \approx 4kT \left( \frac{1}{g_m}+R_F \right)
\end{equation} 
The shape of the PSD in (9) changes with the values of C$\mathrm{_{Par}}$ and C$\mathrm{_F}$ varying with respect to each other. As C$\mathrm{_{Par}}$ increases, the PSD begins to peak at higher frequencies. This high frequency noise can however be attenuated through the subsequent stages since it exceeds the observation bandwidth of the frontend. In this work, using a reset integrator as the following block (section IV-C), the required filtering can be achieved. The boxcar averaging applies a Sinc transfer function to the TIA output as follows:
\begin{figure}[!t]
    \setlength\abovecaptionskip{-0.3\baselineskip}
    \centerline{\includegraphics[width=\columnwidth]{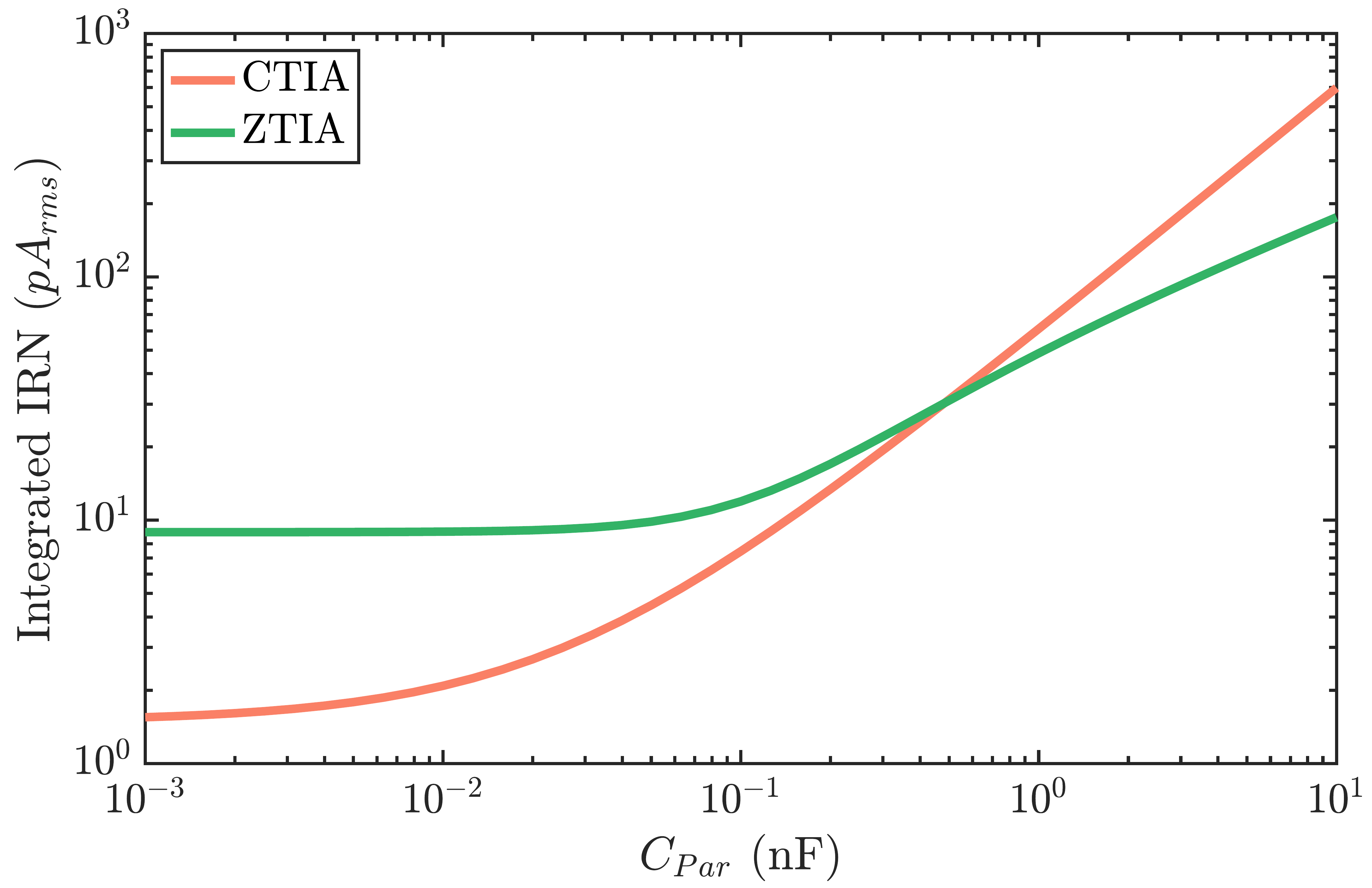}}
    \caption{The input referred standard deviation of a single sample taken in ZTIA and CTIA schemes versus C$\mathrm{_{Par}}$.}
    \label{fig16}
\end{figure}

\begin{equation}
    H_{Sinc}(f) = \frac{T_{Int}}{R_{Int}\cdot C_{Int}} \cdot Sinc(T_{Int}\cdot f)
\end{equation}
Integrating (9) when filtered by (11) will determine the variance of a single sample taken at the ADC output. (9) can also be used to find the noise PSD of the CTIA topology prior to sampling, via setting $R_F \rightarrow \infty$, resulting in:
\begin{equation}
    S_{N_{out_{CTIA}}}(f) = 4kT\frac{1}{g_m}\cdot 1/ \left( 4\pi^2 \frac{C_F^2}{g_m^2} f^2 + \frac{C_F^2}{C_{Par}^2} \right)
\end{equation}
At low frequencies, the OTA input referred voltage noise is boosted by a factor of $C_{Par}/C_F$ at the CTIA output, highlighting the issue of CTIA topology in handling large C$\mathrm{_{Par}}$. In practice, CTIA designs require reset phases inducing reset kT/C noise which will then be removed via CDS. \cite{b15} provides a comprehensive analysis of this scheme. The effective output noise PSD of that scheme can be found as:
\begin{equation}
\begin{gathered}
    S_{N_{CDS}}(f) = 2(\pi f_C T_{Int} - 1) \left( 1 + \frac{C_{Par}}{C_F} \right)^2 \\ \times 4kT \frac{1}{g_m} \cdot Sinc^2(T_{Int}\cdot f)
\end{gathered}
\end{equation}
where $f_C$ is the closed loop bandwidth of the CTIA. By integrating this PSD, one can find the variance of a sample at the CTIA output. Figure 16 shows the variance of samples taken in both CTIA and ZTIA schemes when C$\mathrm{_{Par}}$ changes from 1 pF to 10 nF. For both schemes, similar effective gain, T$\mathrm{_{Int}}$, and g$\mathrm{_m}$ are assumed to deliver a fair comparison. This ensures that the same input can be amplified through the chain over the same observation window and using the same TIA power. Thus, based on the comparison results and since this work aims to acquire current out of a wide range of PDs with potentially very large C$\mathrm{_{Par}}$, the ZTIA scheme is selected in the design.

\section*{Acknowledgment}
The authors thank the NextFlex institute, the CITRIS Seed Award as well as sponsors of Berkeley Wireless Research Center. We also thank TSMC shuttle program for the chip fabrication. Special thanks to Mohammad Meraj Ghanbari for the technical discussions.

\ifCLASSOPTIONcaptionsoff
  \newpage
\fi

\end{document}